\documentclass[twocolumn,showpacs,amsmath,amssymb,prc]{revtex4}

\usepackage{graphicx}
\usepackage{dcolumn}
\usepackage{bm}

\begin{document}


\title{Improved Nearside-Farside Decomposition of Elastic 
Scattering Amplitudes}

\author{R. Anni}
\email{Raimondo.Anni@le.infn.it}
\affiliation{Dipartimento di Fisica dell'Universit\`a and
Istituto Nazionale di Fisica Nucleare, I73100 Lecce, Italy}

\author{J. N. L. Connor}
\email{J.N.L.Connor@Manchester.ac.uk}
\author{C. Noli}
\affiliation{Department of Chemistry, University of Manchester, 
Manchester M13 9PL, United Kingdom}

\date{\today}

\begin{abstract}
A simple technique is described, that provides improved 
nearside-farside (NF) decompositions of elastic scattering 
amplitudes.
The technique, involving the resummation of a Legendre partial
wave series, reduces the importance of unphysical contributions to 
NF subamplitudes, which can arise in more conventional NF decompositions.
Applications are made to a strong absorption model and to a 
$^{16}$O + $^{12}$C optical potential at $E_{\text{lab}} = 132$ MeV.
\end{abstract}

\pacs{24.10.Ht, 25.70.Bc, 34.50.-s, 03.65.Sq}

\maketitle

In heavy-ion, atomic and molecular collisions, an elastic 
differential cross section $\sigma(\theta)$, where $\theta$ is
the scattering angle, is often characterized by a
complicated interference pattern.
This complicated structure makes it difficult to understand the  
physical phenomena involved in the scattering process, as well as 
the links between $\sigma(\theta)$ and the properties of the model 
that describes the phenomenon.

In some cases, semiclassical methods\cite{BRI85} explain the scattering
pattern as the interference between simpler, and slowly varying,
subamplitudes.
If we ignore the complication that in
some angular regions uniform asymptotic techniques are necessary, 
then the semiclassical subamplitudes arise mathematically 
from saddle points or 
poles which account physically 
for the contributions from reflected, refracted or generalized 
diffracted semiclassical trajectories\cite{NUS92}.
The subamplitudes can be conveniently grouped into two types:
those arising from semiclassical trajectories 
which initially move in the same half plane as the detector
(N or nearside trajectories) and those from the opposite half
plane (F or farside trajectories).

Semiclassical methods are not always simple to apply and 
sometimes they have a limited range of applicability. 
Their limitations are determined by the range of validity
for (presently known) asymptotic techniques that provide
an accurate approximation to the original quantum mechanical
problem.

In order to overcome these difficulties, Fuller\cite{FUL75} proposed 
more than 25 years ago, a simple NF decomposition of the
elastic scattering amplitude $f(\theta)$. 
We write $f(\theta)$ in the usual way as a 
partial wave series (PWS)
\begin{equation}
f(\theta )=\frac{1}{2ik} \sum_{l=0}^{\infty }a_{l}P_{l}(\cos \theta ),
\label{ParDev}
\end{equation}
where $k$ is the wavenumber, $P_{l}(\cos \theta )$ is the Legendre
polynomial of degree $l$ and $a_{l}$ is given in terms of the 
scattering matrix element $S_l$ by:
\begin{equation}
a_l= (2 l+ 1) (S_l - 1).
\label{ParAmp}
\end{equation}
We also recall that the PWS in (\ref{ParDev}), considered as a 
distribution, is convergent if $S_l$ is asymptotically 
Coulombic \cite{TAY74}.

The Fuller NF decomposition is realized by splitting $P_l(\cos\theta)$, 
considered  as a standing angular wave, into traveling angular wave 
components
\begin{equation}
P_l(\cos \theta)=Q_l^{(-)}(\cos \theta)+Q_l^{(+)}(\cos \theta),
\label{EulDec}
\end{equation}
where (for $\theta \ne 0, \pi$)
\begin{equation}  
\label{Qpm}
Q_l^{(\mp)}(\cos \theta) = 
\frac {1}{2}[ P_l(\cos \theta) \pm \frac {2i}{\pi}
Q_l(\cos \theta) ],
\end{equation}
with $Q_l(\cos \theta)$ the Legendre function of the second kind of
degree $l $.

Inserting (\ref{EulDec}) into (\ref{ParDev}), splits $f(\theta )$ 
into the sum of two subamplitudes $f^{(\mp )}(\theta )$. 
For  $l \sin \theta \gg 1$, the $Q_{l}^{(\mp )}(\cos \theta )$ behave as
\begin{equation}
Q_{l}^{(\mp )}(\cos \theta )
\sim 
\sqrt{\frac{1}{2\pi \lambda \sin \theta }}
\exp [\mp i(\lambda \theta -\frac{\pi }{4})],  
\label{AsyQpm}
\end{equation}
with $\lambda =l+\frac{1}{2}$.
This asymptotic behavior suggests that the $f^{(\mp )}(\theta )$ should
correspond to NF trajectories respectively appearing in the complete
semiclassical decomposition of $f(\theta )$.

However the NF decomposition is in general less satisfactory than the
full semiclassical one.
For example, if two or more N or F semiclassical trajectories contribute to
the same $\theta$, interference effects may appear in the N
or F cross sections.
The Fuller NF decomposition has, however, the merit of being simple 
and, although inspired by the semiclassical theories, it
uses only quantities calculated within the exact quantum mechanical
treatment.
The NF method therefore bypasses problems associated with
the applicability and validity of semiclassical 
techniques. 

The physical meaning attributed to the $f^{(\mp )}(\theta )$ is 
implicitly based on the (unproven) assumption that 
it is possible to perform on the PWS, written in terms of the
$Q_l^{(\mp )}(\cos \theta)$, the same manipulations which are
used in the complete semiclassical decomposition of $f(\theta)$.
These manipulations are essentially path deformations 
in $\lambda$ of the integrals into which (\ref{ParDev}) 
can be transformed, using either the Poisson summation formula
or the Watson transformation.
The consequences of these path deformations depend on the
properties of the 
terms in the PWS when they are continued to real or complex
values of $\lambda $ from the initial half integer $\lambda$
values.
The splitting of $P_{l}(\cos \theta )$ into $Q_{l}^{(\mp)}(\cos \theta )$ 
modifies these properties and can cause the appearance 
of unphysical contributions in the $f^{(\mp )}(\theta )$ which cancel out
in $f(\theta)$.

In spite of these possible limitations, the Fuller NF decomposition is
widely used. In many cases \cite{HUS84,BRA97}, it decomposes 
$f(\theta)$ into simpler subamplitudes which are free from the unphysical
contributions that can arise from the above mathematical difficulties.
However for a few examples, the NF subamplitudes can be directly
compared with the corresponding semiclassical results and it is found
that the Fuller and semiclassical decompositions predict different results.
One classic example is pure Coulomb scattering. 
For repulsive Coulomb potentials only a N contribution is
expected semiclassically (\cite{BRI85}, p. 56), whereas the Fuller 
NF decomposition yields also a F contribution \cite{FUL75}.
Another example is the angular distribution for a strong
absorption model (SAM) with a two parameter ($\Lambda$ 
and $\Delta$) symmetric $S$ matrix element and Fermi-like 
form factors \cite{HAT89}.
For a fixed value of the cut-off parameter $\Lambda$ and for 
a sufficiently large value of the diffuseness parameter $\Delta$, 
the Fuller NF cross sections agree with the semiclassical
results only up to a certain value of $\theta$, which decreases with 
increasing $\Delta$.

Fortunately, the Fuller NF subamplitudes contain information
which allows one to recognize the unphysical nature of the 
undesired contributions.
Suppose $f^{(+)}(\theta)$, or $f^{(-)}(\theta)$, contains 
a single contribution from a stationary phase point at 
$\lambda(\theta$). Then the derivative with respect to 
$\theta$ of the phase of $f^{(+)}(\theta)$, or $f^{(-)}(\theta)$, 
is equal to $\lambda(\theta)$, or -$\lambda(\theta)$ respectively; it 
depends on $\theta$ (\cite{BRI85}, p. 57).
Following Fuller we will call this derivative the {\it Local Angular 
Momentum} (LAM) for the N (or F) subamplitude.
Only for certain generalized diffracted trajectories is the LAM  
expected to be constant, equal to the angular momentum 
of the incoming particle responsible for 
the diffraction.
In the semiclassical regime, this constant value is expected 
to be large. Because of this, if we observe in a certain $\theta$ 
range that LAM $\approx$ 0, this can be considered the 
signature of the unphysical nature of the N or F subamplitudes 
in that range of $\theta$.
This occurs for the LAM of the Fuller Coulomb F 
subamplitude, and for  the NF subamplitudes of the SAM in 
the angular region where the NF cross sections
differ from the semiclassical results.
In both cases this decoupling of $\theta$ from LAM  
suggests the unphysical nature of the subamplitudes.
Thus an analysis of the LAM can avoid misleading interpretations 
of cross sections obtained from the Fuller NF decomposition. 

However the problem of obtaining more satisfactory NF decompositions 
remains open.
A possible solution to the problem was proposed by Hatchell 
\cite{HAT89}, who used a modified NF decomposition.
The modifications consisted of, first, in writing $f(\theta)$ 
in the resummed form ($\theta \ne 0$) 
\begin{equation}
\label{JenDev}
f(\theta )=\frac{1}{2 i k} \frac{1}{(1-\cos \theta )^{r}}
\sum_{l=0}^{\infty} a_{l}^{(r)}P_{l}(\cos \theta ),
\end{equation}
$r =1,2,\ldots$, and, second, in using a different splitting for 
the Legendre polynomials into traveling waves. 

The use of the resummed form (\ref{JenDev}) for $f(\theta)$ was 
originally proposed \cite{YEN54} by Yennie, Ravenhall, and Wilson 
(YRW) to speed up the convergence of the PWS for
high-energy electron-nucleus scattering.
Equation (\ref{JenDev}) is an exact resummation formula, of order $r$,
which is derived from the recurrence relation for Legendre polynomials.
The YRW resummation formula can be derived by iterating $r$ times,
starting from $a_{l}^{(0)}=a_{l}$, the 
resummation identity 
\begin{equation}
\label{JenIde}
\sum_{l=0}^{\infty} a_{l}^{(i-1)}P_{l}(\cos \theta)
=
\frac{1}{1 - \cos \theta}
\sum_{l=0}^{\infty} a_{l}^{(i)}P_{l}(\cos \theta),
\end{equation}
where
\begin{equation}
a_{l}^{(i)}=-\frac{l}{2l-1} a_{l-1}^{(i-1)}+a_{l}^{(i-1)}-\frac{l+1}{2l+3}%
a_{l+1}^{(i-1)},
\label{JenRec}
\end{equation}
with $a_{-1}^{(i-1)}= 0$. 

Note that $f(\theta)$ does not depend on $r$, unlike the Fuller
NF subamplitudes which do depend on the value of 
$r$ used.
This is a consequence of the property $l Q_{l-1} (\cos \theta) \rightarrow 1$ 
as $l \rightarrow 0$ \cite{ANN81}.
In the Hatchell approach, the dependence on $r$ arises because 
the functions used in place of the 
$Q_l(\cos \theta)$ obey a recurrence (inhomogeneous) relation 
different from that for $Q_l(\cos \theta)$.
It is worth-while to note that (\ref{JenDev}) for $r \ge 1$ 
allows one to drop, for $\theta > 0$, the 1 appearing in 
the term $S_l - 1$ in  (\ref{ParAmp}).
Furthermore for  $r \ge 1$, (\ref{JenDev}) produces a 
convergent PWS even when 
$S_l$ is asymptotically Coulombic \cite{GOO80}.

Using his method, Hatchell has shown \cite{HAT89} that the 
unphysical contributions to the SAM NF cross sections 
systematically decrease on increasing $r$.
More recently \cite{HOL99a}, it was shown that, using the Fuller
$Q_{l}^{(\mp)}(\cos \theta)$ functions in (\ref{JenDev}), 
gives even better results.
The superiority of the $Q_{l}^{(\mp)}(\cos \theta)$ functions 
seems to be connected with the greater rapidity with which the 
$Q_{l}^{(\mp)}(\cos \theta )$ approach their asymptotic behavior 
(\ref{AsyQpm}) \cite{McC95,HOL99b}, compared to
the Hatchell NF functions.

The success of using (\ref{JenDev}) before
applying the NF decomposition depends upon
the properties of the $a_l^{(r)}$.
For the SAM, the contributions from low $l$ values rapidly
decrease\cite{HOL99a} with increasing  $r$.
As a result, the most important partial waves move to 
higher values of $l$, where a semiclassical description 
is physically more reasonable. 

However, in some cases, (\ref{JenDev}) acts in the opposite direction,
by enhancing the undesired unphysical contributions to the NF
subamplitudes.
We have found that this happens, for example, 
for pure Coulomb scattering, for scattering by an impenetrable 
sphere, and for the SAM (see \cite{NOLXX}) when the cross section 
is calculated at an angle $\pi -\theta$, using the property 
$P_l[\cos (\pi-\theta)]=(-1)^l P_l(\cos \theta)$.

One possible solution to this intriguing puzzle is to regard 
(\ref{JenIde}) as a particular case of the modified
resummation identity \cite{WHI01}
\begin{equation}
\label{JenIdeGen}
\sum_{l=0}^{\infty} a_{l}^{(i-1)}P_{l}(\cos \theta)
=
\frac{1}{\alpha_{i} + \beta_{i} \cos \theta}
\sum_{l=0}^{\infty} a_{l}^{(i)}P_{l}(\cos \theta),
\end{equation}
with $\alpha_i+ \beta_i \cos \theta \ne 0$ and
\begin{equation}
\label{JenRecGen}
a_{l}^{(i)}= \beta_{i} \frac{l}{2 l-1} a_{l-1}^{(i-1)}
+ \alpha_{i} a_{l}^{(i-1)}
+ \beta_{i} \frac{l+1}{2 l+3} a_{l+1}^{(i-1)}.
\end{equation}
For $\alpha_i, \beta_i \ne 0$, the  r.h.s. of
(\ref{JenIdeGen}) depends only on the ratio $\beta_{i}/\alpha_{i}$.
Thus, without loss of generality, we can assume $\alpha_{i}=1$
for all $i$.
By iterating (\ref{JenIdeGen}) $r$ times, we can write $f(\theta)$
in the modified resummed form
\begin{equation}
\label{JenDevGen}
f(\theta )=\frac{1}{2 i k}
\left ( 
\prod_{i=1}^{r} \frac{1}{1+\beta_i \cos \theta }
\right )
\sum_{l=0}^{\infty} a_{l}^{(r)}P_{l}(\cos \theta ),
\end{equation}
$r = 1, 2, \ldots \, $. 
The YRW resummation formula (\ref{JenDev}) is  
obtained with $\beta_{1} = \beta_{2}= \ldots = \beta_{r} = -1$.

The resummation identity (\ref{JenIdeGen}) is a particular case of a
more general one \cite{WHI01}, which uses a basis set
of reduced rotation matrix elements; this gives 
the amplitude for more general scattering processes
than those described by (\ref{ParDev}).
For these general PWS, a Fuller-like NF decomposition
can be introduced \cite{ABI99,SOK99,McC01} which
allows the scattering amplitude to be split into 
NF subamplitudes. 
In some cases, the NF cross 
sections contained unexpected (unphysical) oscillations \cite{WHI01}, 
which are enhanced if the generalization of 
(\ref{JenDev}) is used, but which disappear for an appropriate choice 
of the $\beta$-parameter in the generalization of (\ref{JenDevGen}).

The considerable successes achieved by the original Fuller 
NF decomposition suggests that the modified
resummed form (\ref{JenDevGen}) be used to diminish 
unphysical contaminations to the NF subamplitudes
when they are present.
To do this, we must give a practical rule to fix the values
of the $\beta$-parameters.
In Refs. \cite{NOLXX,WHI01} it was proposed to select the 
value of $\beta \equiv \beta_1 =\ldots=\beta_r$ so that 
$(1+\beta \cos \theta)^{-r}$  {\it approximately mimics 
the shape} of the angular distribution.
The shape of the cross section can however be very
different from that given by 
$(1+\beta \cos \theta)^{-r}$.
It is therefore desirable to test a different recipe,
based on a simple rule.
The quantitative recipe proposed here is inspired by the 
observation that the modified resummation formulas produce 
a more physical NF decomposition by reducing the contribution 
from the low $l$ values in the resummed PWS.
This suggests we select the $\beta_1, \beta_2, \ldots, 
\beta_r$ in $r$ repeated applications of (\ref{JenIdeGen}), 
so as to eliminate as many low partial waves as possible 
from the final PWS in  (\ref{JenDevGen}).
The transformation from $\{a_l^{(i-1)}\}$ to $\{a_l^{(i)}\}$
is linear tridiagonal, with coefficients linear in
$\beta_{i}$, which means application of $r$ successive 
resummations allows one to equate to zero the leading $r$ 
coefficients $a_l^{(r)}$, with $l=0,1,\ldots,r-1$, by solving 
a system of $r$ equations of degree $r$ in
the parameters $\beta_1, \beta_2, \ldots, \beta_{r}$.
We will call the resummation defined in this way an
{\it improved resummation of order $r$}.

It is straightforward to show that the improved 
resummation of order 1 is obtained by choosing
\begin{equation}
\beta_1=-3 a_0 / a_1,
\label{OneOpt}
\end{equation}
while the improved resummation of order 2 is given by
\begin{equation}
\beta_{1,2}=(B \pm \sqrt{B^2-4A}) /2 \, ,
\label{TwoOpt}
\end{equation}
with $A$ and $B$ solutions of the linear equations
\begin{equation}
\left \{
\begin{array}{lcr}
(\frac{1}{3} a_0+\frac{2}{15} a_2) A &+  &\frac{1}{3} a_1 B =-a_0\,
\\
\\
(\frac{3}{5} a_1+\frac{6}{35} a_3) A &+ &(a_0+\frac{2}{5} a_2) B =-a_1.
\end{array}
\right.
\label{SYS}
\end{equation}
Higher order improved resummations require the solution of more
complicated systems of equations.
 
In all the cases we have analyzed using $r \leq 2$, we find that the
improved resummations considerably reduce the width of the 
angular regions in which the Fuller NF cross sections exhibit 
unphysical behavior.
In these analyses, we have used $S_l$ from simple parametrizations 
as well as from some of the optical potentials currently employed 
to describe light heavy-ion scattering.
We show below our results for two particular examples.
The first example is a SAM (Fig. 1), whilst the
second example is the $^{16}$O + $^{12}$C collision, at
$E_{\text {lab}}$ = 132 MeV, using the WS1 optical potential of
Ref.\cite{OGL00} (Fig. 2).
For both these cases we have dropped the 1 in the term  $S_l -1$ in Eq.
(\ref{ParAmp}). 
The calculations were performed applying: {\it first},
an improved resummation of order $r = 0, 1, 2$, with $r = 0$ meaning
no resummation, {\it second}, the Fuller NF decomposition (\ref{EulDec}), 
(\ref{Qpm}), and {\it third} a YRW resummation of the NF subamplitudes
using the extension of (\ref{JenDev}) 
to the linear combination (\ref{Qpm}) of integer degree Legendre 
functions of the first and second kinds \cite{ANN81}.
This latter resummation ensures the convergence of the final NF PWS.
The results obtained from these three steps will be indicated
by the notation $\text{R} =0_{\text Y}, 1_{\text Y}, 2_{\text Y}$.

For the SAM we have chosen the parameters to be $\Lambda=10$ and 
$\Delta=2$. In Fig. 1 we have plotted the dimensionless 
quantity $4 k^2 \sigma(\theta) \sin \theta$ since the corresponding
NF semiclassical quantities are expected to have a pure exponential 
slope \cite{HAT89}.
Furthermore, because the $S_l$ are real, $f(\theta)$ has a
constant phase (and its phase derivative is of no interest), 
and the $f^{(\mp)}(\theta)$ have the same moduli but opposite phases. 
Thus we need only show the N, or F, LAM and similarly for the cross 
sections.

Figure 1 also shows the results obtained on introducing 
the NF decomposition directly into (\ref{ParDev}) and without 
dropping the 1 in (\ref{ParAmp}). The NF subamplitudes obtained in 
this way are rapidly convergent and no final YRW 
resummation is needed. These results are indicated by
the notation $\text{R} = 0$. 

For the original Fuller NF method, $\text{R} = 0$, and for the case 
$\text{R} = 1_{\text Y}$, we have plotted the F cross sections and LAMs 
(dashed curves);
for the cases $\text{R} = 0_{\text Y}$ and $\text{R} = 2_{\text Y}$ the N quantities 
(continuous curves) are displayed.  
\begin{figure}
\centering
\includegraphics[width = 8.6 cm]{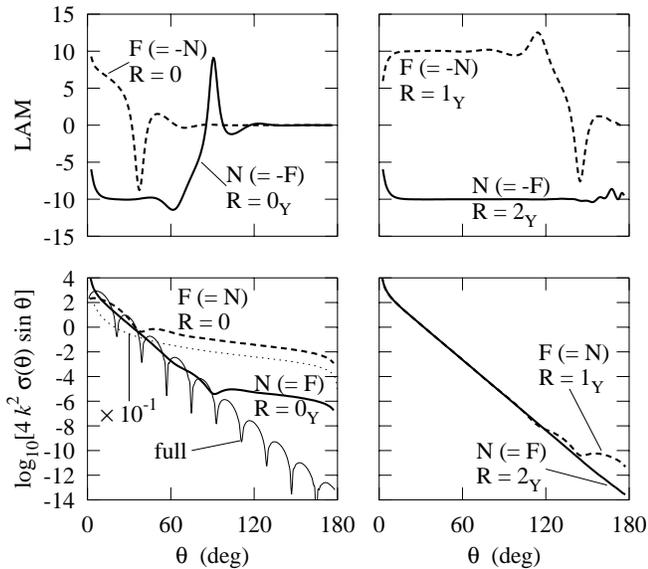}
\caption{\label{fig:Fig01} 
Strong absorption model N (continuous lines) and F
(dashed lines) cross sections (lower panels)
and LAM (upper panels) calculated using the 
$\text{R} = 0, 0_{\text Y}, 1_{\text Y}, 2_{\text Y}$ NF decompositions.
The thin curve shows the full cross section. The thin dotted
curve shows the F (= N) cross section (displaced downward by one unit) for the 
unphysical amplitude $f^{(+)}_\delta(\theta)$.}
\end{figure}
Figure 1 shows that for $\text{R} = 0$ the unphysical contributions 
dominate the F (= N) cross section over most of the angular range.
The expected exponential behavior is not present in the F cross section 
curve and the F (= -N) LAM $\approx 0$ for $\theta \gtrsim 60^\circ$. 
However, at smaller angles, oscillations in the F LAM curve 
indicate that another contribution is present which interferes 
with the unphysical one. 
This behavior does not support the conjecture that the F LAM of this other 
contribution has the semiclassical value $\Lambda$.
The major part of the unphysical contribution, 
which dominates the F subamplitude, is 
$f^{(+)}_\delta(\theta) = -[2 \pi k (1-\cos \theta)]^{-1}$.
This is the F component of $f_\delta(\theta) = i \delta(1-\cos \theta)/k$, obtained
by dropping $S_l$ in the term $S_l -1$ in (\ref{ParAmp}).
%
%
The cross section of $f^{(+)}_\delta(\theta)$ is shown, downward shifted by 
one vertical unit, by the thin dotted curve in Fig. 1. 

The $\text{R} =0_{\text Y}$ method provides more satisfactory results, 
which are rather good at forward angles. 
With the exclusion of a small region around $\theta = 0^\circ$,
where (\ref{AsyQpm}) does not hold, and up to $\theta \approx 50^\circ$, 
the N (= -F) LAM agrees closely  with the expected semiclassical value 
of $-\Lambda$ and the N (= F) cross section curve follows the expected 
exponential behavior. 
For $\theta \gtrsim 120^\circ$, the N cross section is still dominated
by an unphysical contribution.
At intermediate angles, $50^\circ \lesssim \theta \lesssim 120^\circ$, 
interference oscillations appear both in the N cross section and in 
the N LAM curve.
It is interesting to note that the LAM is more sensitive
to interference effects than is the cross section. Also, 
in the interference region, one cannot attach the meaning
of a {\it local angular momentum} to the subamplitude
phase derivative.
In our case, in this interference region, the N LAM curve oscillates
around the expected semiclassical value of $-\Lambda$ in the 
region, $50^\circ \lesssim \theta \lesssim 80^\circ$, 
 where the true semiclassical component dominates the 
N subamplitude, and around the unphysical value of 0 at other angles.

The effectiveness of the improved resummation procedure
is evident in the right panels of Fig. 1. 
Using the $\text{R} =1_{\text Y}$ method ( for which $\beta_1=-0.800$) 
the F (= -N) LAM and the F (= N) cross section are in agreement with 
the semiclassical results up to $\theta \approx 120^\circ$. 
For $\text{R} =2_{\text Y}$ (which has $\beta_{1, 2}=-0.879 \pm 0.076 \, i$)
the agreement covers almost the whole angular range. 
The small irregular oscillations appearing at large $\theta$ for the 
N LAM curve, with $\text{R} =2_{\text Y}$, probably arise from the precision limitations 
(64 bit floating point representation) of the calculations.
\begin{figure}
\centering
\includegraphics[width = 8.6 cm]{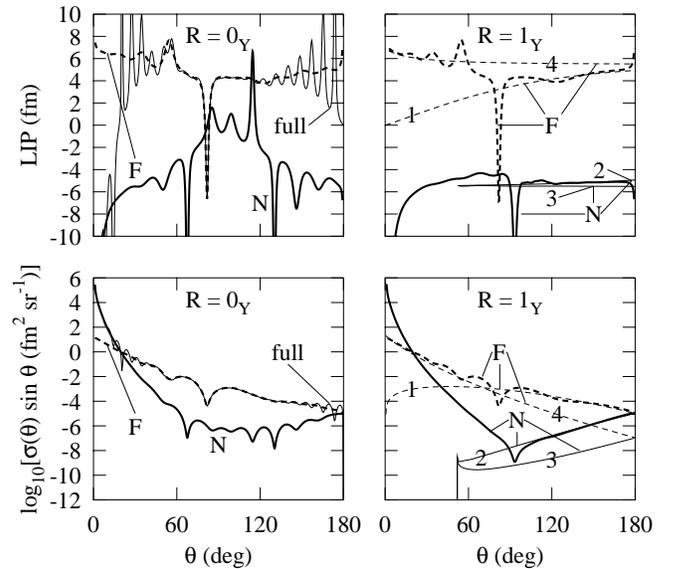}
\caption{\label{fig:Fig02} 
Optical potential model N (continuous lines) and F
(dashed lines) cross sections (lower panels)
and LIP (upper panels) calculated using the $\text{R} =0_{\text Y}$ and 
$\text{R} =1_{\text Y}$ NF decompositions.
The thin curves show the cross section and LIP obtained using the full quantum amplitude
in the left panels, and the N (continuous) and F (dashed) cross sections
and LIP using classical mechanics in the right panels. The indices in the right
panels identify the curves corresponding to different branches of the classical 
deflection function.}
\end{figure}

Figure 2  shows our results for the optical potential. 
In the upper panels we display LAM/$k$, which we call 
the {\it Local Impact Parameter} (LIP), and in the lower 
panels a Log plot of $\sigma (\theta) \sin \theta$.
The left panels show the results for the usual Fuller NF 
decomposition, $\text{R} =0_{\text Y}$.
The thin continuous lines, in the left panels, show the  
cross section and LIP for the full amplitude. 

The behavior of the NF LIP curves is mostly simpler than that of 
the full amplitude LIP. 
At $\theta \approx 90^\circ$, the $\text{R} =0_{\text Y}$ N LIP curve 
oscillates around 0, indicating the possible dominance of an 
unphysical contribution. 
This contribution also appears to be responsible for oscillations
in the N LIP curve around other values (different from 0),    
and for oscillations in the N cross section for $\theta \gtrsim 30^\circ$.
These oscillations are absent in the N curves in the 
right panels where the results for $\text{R} =1_{\text Y}$ are shown
($\beta_1=-0.999-0.099 \, i$). 
Both the N cross section and N LIP curves for $\text{R} =1_{\text Y}$ are 
considerably simpler than those obtained using $\text{R} =0_{\text Y}$, while
the F curves are essentially the same; an exception is the less
oscillatory F LIP for $\theta \gtrsim 120^\circ$.
This indicates that, apart from $\theta \gtrsim 120^\circ$, the 
unphysical contribution for $\text{R} =0_{\text Y}$ has a modulus 
which is much smaller then that of the F semiclassical subamplitude.    
We have also applied the improved resummation $\text{R} =2_{\text Y}$. The results
are practically the same as those obtained using $\text{R} =1_{\text Y}$ and are 
not shown.

The cleaning by the $\text{R} =1_{\text Y}$ procedure 
of the original $\text{R} =0_{\text Y}$ NF subamplitudes is 
impressive and allows a clear 
identification, in the NF cross sections at $\theta \gtrsim 120^\circ$,
of the dominance of semiclassical trajectories refracted from the nuclear 
part of the interaction.
In the right panels of Fig. 2, this interpretation is confirmed 
by the agreement, for $\theta \gtrsim 120^\circ$, between the NF 
curves and the corresponding classical mechanical results 
(thin lines 1 and 2).
The thin lines show, in the upper panel, different NF branches of the
the impact parameter and their dependence on $\theta$ 
(with appropriate signs) using only the real part of the complete 
interaction. 
In the lower panel we show the classical contributions to the 
cross section from these branches, in which we have included in the 
usual simple way (\cite{BRI85}, p. 49) the 
absorptive effects of the imaginary part of the optical potential.
   
Our new resummation NF procedure clearly improves the original Fuller 
NF decomposition, as is evident in the examples presented here.
On the one hand, our results confirm the importance of 
NF decompositions for gaining insight into the properties of
the subamplitudes responsible for complicated structures
in cross sections. 
On the other hand, they confirm the empirical origin of NF 
decompositions and suggest caution in the interpretation of  
results obtained from NF techniques.
However, different NF decompositions can be used to check what parts
of the resulting NF subamplitudes
are independent of the particular technique used.
Only properties stable with respect to different NF
decompositions, can be considered as manifestation of some 
physical phenomenon.
In addition, we have shown that it is desirable to investigate
the behavior of the LAM. 
This quantity is more sensitive to interference effects than
are the NF cross sections, and a null value (or an oscillatory
behavior around zero) of the LAM in a certain angular range
indicates an unphysical contribution.

\begin{acknowledgments}
Support of this research by a PRIN MIUR research grant (I), the
Engeneering and Physical Sciences Research Council (UK) and INTAS (EU)
is gratefully acknowledged.
\end{acknowledgments}

\bibliography{NF_side}

\end{document}